\documentclass[%
 reprint,
 amsmath,amssymb,
 aps,
]{revtex4-2}

\usepackage{graphicx}% Include figure files
\usepackage{dcolumn}% Align table columns on decimal point
\usepackage{bm}% bold math

\usepackage[T1]{fontenc} 
\begin{document}

\preprint{APS/123-QED}

\title{Quantitative and dark field ghost imaging with ultraviolet light}% Force line breaks with \\

\author{Jiaqi Song\textsuperscript{1}}
\author{Baolei Liu\textsuperscript{1}}
 \email{liubaolei@buaa.edu.cn}
\author{Yao Wang\textsuperscript{1}}
\author{Chaohao Chen\textsuperscript{2}}
\author{Xuchen Shan\textsuperscript{1}}
\author{\\Xiaolan Zhong\textsuperscript{1}}
 \email{zhongxl@buaa.edu.cn}
\author{Ling-An Wu\textsuperscript{3}}
\author{Fan Wang\textsuperscript{1}}

\affiliation{%
 \textsuperscript{1}School of Physics, Beihang University, Beijing, 102206, China\\
\textsuperscript{2}Australian Research Council Centre of Excellence for Transformative Meta-Optical Systems, Department of Electronic Materials Engineering, Research School of Physics, The Australian National University, Canberra, ACT 2600, Australia\\
 \textsuperscript{3}Institute of Physics, Chinese Academy of Sciences, Beijing 100190, China
}%

\date{July 17, 2023}% It is always \today, today,

\begin{abstract}
Ultraviolet (UV) imaging enables a diverse array of applications, such as material composition analysis, biological fluorescence imaging, and detecting defects in semiconductor manufacturing. However, scientific-grade UV cameras with high quantum efficiency are expensive and include a complex thermoelectric cooling system. Here, we demonstrate a UV computational ghost imaging (UV-CGI) method to provide a cost-effective UV imaging and detection strategy. By applying spatial-temporal illumination patterns and using a 325 nm laser source, a single-pixel detector is enough to reconstruct the images of objects. To demonstrate its capability for quantitative detection, we use UV-CGI to distinguish four UV-sensitive sunscreen areas with different densities on a sample. Furthermore, we demonstrate dark field UV-CGI in both transmission and reflection schemes. By only collecting the scattered light from objects, we can detect the edges of pure phase objects and small scratches on a compact disc. Our results showcase a feasible low-cost solution for non-destructive UV imaging and detection. By combining it with other imaging techniques, such as hyperspectral imaging or time-resolved imaging, a compact and versatile UV computational imaging platform may be realized for future applications.
\end{abstract}

%\keywords{Suggested keywords}%Use show keys class option if keyword
\maketitle

%\tableofcontents

\section{Introduction} %%%%%%%%%%%% Introduction
Ghost imaging (GI), an alternative imaging technique that uses only a single-pixel detector, has been widely studied in both quantum and classical optics \cite{1pittman1995optical,2padgett2017introduction}. Traditional GI uses a beam splitter to divide random light fields into two light paths \cite{3zhang2005correlated}; one path uses an arrayed detector to record the spatial distribution of the random light field, and the other uses a single-pixel detector (SPD) to collect the intensity of the transmitted or reflected light from an object in that beam. The object’s image can be reconstructed via the second-order correlation function between the measurement of these two detectors. Computational ghost imaging (CGI), as a modification of GI, removes the need for the arrayed detection of the light field by adopting a spatial light modulator (SLM) to generate a sequence of programmable patterns to illuminate the object \cite{4shapiro2008computational,5bromberg2009ghost}. This is closely related to the scheme of single-pixel cameras, which use the SLM to modulate the field at the image plane and an SPD to measure the light intensities after the modulation \cite{6duarte2008single,7edgar2019principles}. One advantage of GI/CGI is that it can obtain an image with many fewer measurements than the total number of image pixels, a strategy also known as compressive sampling or sub-sampling \cite{6duarte2008single,8gibson2020single}. Due to its easy implementation, GI/CGI has been extended into various applications, such as LiDAR \cite{9zhao2012ghost,10sun2016single}, cytometry \cite{11ota2018ghost}, hyperspectral imaging \cite{12olivieri2020hyperspectral,13yi2020hadamard,14jin2017hyperspectral}, phase imaging \cite{15liu2018single,16zhao2023single}, and so on. Another advantage of GI/CGI is that it can be achieved with a wide range of electromagnetic waves since different kinds of spatial modulators can be used, for example, rotating ground-glass diffusers \cite{9zhao2012ghost,17chen2009lensless,18liu2021single,41dou2020dark}, liquid crystal SLMs \cite{19li2022dual}, digital micromirror devices (DMDs) \cite{20liu2021self,21wang2023dual}, or coded masks \cite{22hahamovich2021single,23jiang2021single}. To date, GI/CGI has been realized with entangled photon pairs \cite{1pittman1995optical,33bornman2019ghost}, visible light \cite{27ye2020simultaneous,28wang2020all}, near-infrared \cite{29gibson2017real}, mid-infrared \cite{30wang2023mid}, and terahertz wavelengths \cite{19li2022dual,31watts2014terahertz,32stantchev2017compressed}, X-rays \cite{24pelliccia2016experimental,25zhang2018tabletop,26yu2016fourier,he2020high}, atoms \cite{34khakimov2016ghost}, electrons \cite{35li2018electron} and neutrons \cite{kingston2020neutron,he2021single}. However, GI/CGI has rarely been studied in the ultraviolet (UV) band. Recently, single-pixel cameras in the UV band were employed to capture flame chemiluminescence \cite{36zhang2019demonstration} and identify different transparent objects with a single-photon avalanche diode \cite{46ye2023ultraviolet}. Many UV-sensitive applications with active illumination need to be further studied.

Generally, UV imaging can be divided into two types: UV-fluorescence imaging and UV-reflectance imaging. In the former, objects absorb UV light and emit fluorescence of longer wavelengths, such as in real-time non-destructive monitoring of food freshness \cite{38zhuang2022uv}. In the latter, a UV camera is used to collect the light reflected from the objects. UV reflectance imaging can detect hidden damages that are not sensitive to visible light / human eyes \cite{king2018use,patel2019potential}. Due to its short wavelength, UV light is more easily scattered by smaller features which are not obvious at longer wavelengths \cite{39shaw2009deep,40ren2022state}. However, high-performance UV cameras are expensive, which limits their usage in practical applications.

In this study, we report the applications of UV-CGI in quantitative detection and dark-field imaging. First, we used samples containing UV-sensitive sunscreen. Second, a sample was formed by dispersing different amounts of sunscreen on a piece of paper so that the different grayscale values of the reconstructed image revealed the different thicknesses of sunscreen. The lower the grayscale value, the higher the dosage of sunscreen due to the higher absorption under UV light irradiation. Furthermore, we propose a dark-field UV-CGI for detecting phase objects and slight damage on a compact disc (CD). By only detecting the scattered light from the samples, we can retrieve the edges of the phase objects and images of the defects. Our technique broadens the application possibilities of GI in the UV range and provides a scalable low-cost UV imaging method, which can be further combined with hyperspectral and time-resolved imaging and be integrated into miniature optical systems.

\section{Theory}
\subsection{The principles of ghost imaging}
In GI, an object $O(x_0,y_0)$ is illuminated sequentially by $N$ illumination patterns $P$, the intensities $I$ of the corresponding total reflected or transmitted light, which is measured by an SPD, can be expressed as:
\begin{equation}
I_i=\int P_i(x_0,y_0) \cdot O(x_0,y_0)dx_0dy_0,
\end{equation}
where $i=1,2,3 \cdots,N$ is the index of the pattern, and $x_0$ and $y_0$ are spatial coordinates. To retrieve the image, a traditional correlation algorithm can be used, such as \cite{5bromberg2009ghost}:
\begin{equation}
O(x_0,y_0)=\textless IP(x_0,y_0) \textgreater - \textless I \textgreater \textless P(x_0,y_0) \textgreater,
\end{equation}
where $\textless \cdots \textgreater $ is the ensemble average over the distribution of the patterns. By applying compressive sensing algorithms, high-quality images can be retrieved with a low sampling ratio.

\subsection{The principles of dark-field UV-CGI}
We further develop dark-field UV-CGI that has advantages in imaging phase objects and small scattering features such as scratches on a CD. Compared with traditional GI/CGI, dark-field CGI only collects the scattered light from objects, instead of the total intensity $I_i$, by using a beam stop to block the directly reflected/transmitted light (the detailed experimental setup is presented in Figs. 4 and 5 and the following section). After the pattern illuminates the object, the total electric field amplitude $E_{total} $ at the object plane can be briefly expressed as follows:
\begin{equation}
E_{total}=m(x_0,y_0)=E_r(x_0,y_0)+E_t(x_0,y_0)+E_s(x_0,y_0),
\end{equation}
where $E_r(x_0,y_0)=O_r(x_0,y_0)\cdot E_i(x_0,y_0)$ is the reflected wave, and $E_t(x_0,y_0)=O_t(x_0,y_0)\cdot E_i(x_0,y_0)$ is the transmitted wave, and $E_s(x_0,y_0)$ is the scattered wave from the object, and $O_r(x_0,y_0)$ and $O_t(x_0,y_0)$ are the coefficients of reflection and transmission, respectively. In the experiment, the SPD only collects the scattered wave ($E_s(x_0,y_0)$), which can be achieved by blocking the directly reflected/transmitted light in front of the detector, as shown in Figs. 4 and 5. Thus, after the light passes through the beam block, the reflected wave or the transmitted wave is blocked such that $O_r(x_0,y_0)=O_t(x_0,y_0)=0$. The light field amplitude distribution at the plane of the beam block can be expressed as follows:
\begin{equation}
\begin{aligned}
E_B(x,y) &= m(x,y) \int E_s(x_0,y_0) \\
& \cdot exp \{ \frac{ik}{2r} [(x-x_0)^2+(y-y_0)^2] \}dx_0dy_0,\\
\end{aligned}
\end{equation}
where $x$ and $y$ are the spatial positions of the beam block, $r$ is the distance between the object and the beam block, and $m(x,y)$ is the pupil function of the beam block. When light is incident on an object, a portion of the light will scatter depending on the geometry of the object. For pure phase objects, as the phase gradient increases, scattering will also increase which can be used to reveal the edge of the objects. Therefore, for dark-field GI, we can observe the edges by collecting the partially scattered light. The detected intensity $I_{DF}$ can be expressed as follows: 
\begin{equation}
\begin{aligned}
I_{DF} &= \int \Big\{ \int E_B(x,y) \\
& \cdot exp [\frac{ik}{2r} ((x^{'}-x)^2+(y^{'}-y)^2)]dxdy \Big\}^2dx^{'} dy^{'},\\
\end{aligned}
\end{equation}
where $x^{'}$ and $y^{'}$ are the spatial positions of the SPD, and $r^{'}$ is the distance from the beam block to the SPD, and $E_s(x,y)$ is the amplitude distribution of the scattered light after passing through the beam block. The edge of the phase object can be considered as $O^{''}(x_0,y_0)$, which indicates the alteration of the phase object $O(x_0,y_0)$. In the setup, the SPD is close to the beam block. Thus, $I_{DF} \approx \int |E_{B}(x,y)|^2dxdy$. The edge of the phase object $O^{''}(x_0,y_0)$ can be retrieved by the following equation:
\begin{equation}
O^{''}(x_0,y_0)=\textless I_{DF}P(x_0,y_0) \textgreater - \textless I_{DF} \textgreater \textless P(x_0,y_0) \textgreater.
\end{equation}

Note that not all the scattered light can be collected by the SPD, as this depends on the sensitive area of the beam stop and the SPD. A smaller beam block area can allow more scattered light to be collected but cannot be smaller than the cross-sectional area of the incident beam. Moreover, although a larger detector area allows for collecting more scattered light, it may increase the equipment cost. In section 3.3 we show that UV-CGI can detect the edges of pure phase objects and small scratches on a CD.
\begin{figure*}[!!t]
\centering
\includegraphics[width=1\textwidth,height=0.402\textwidth]{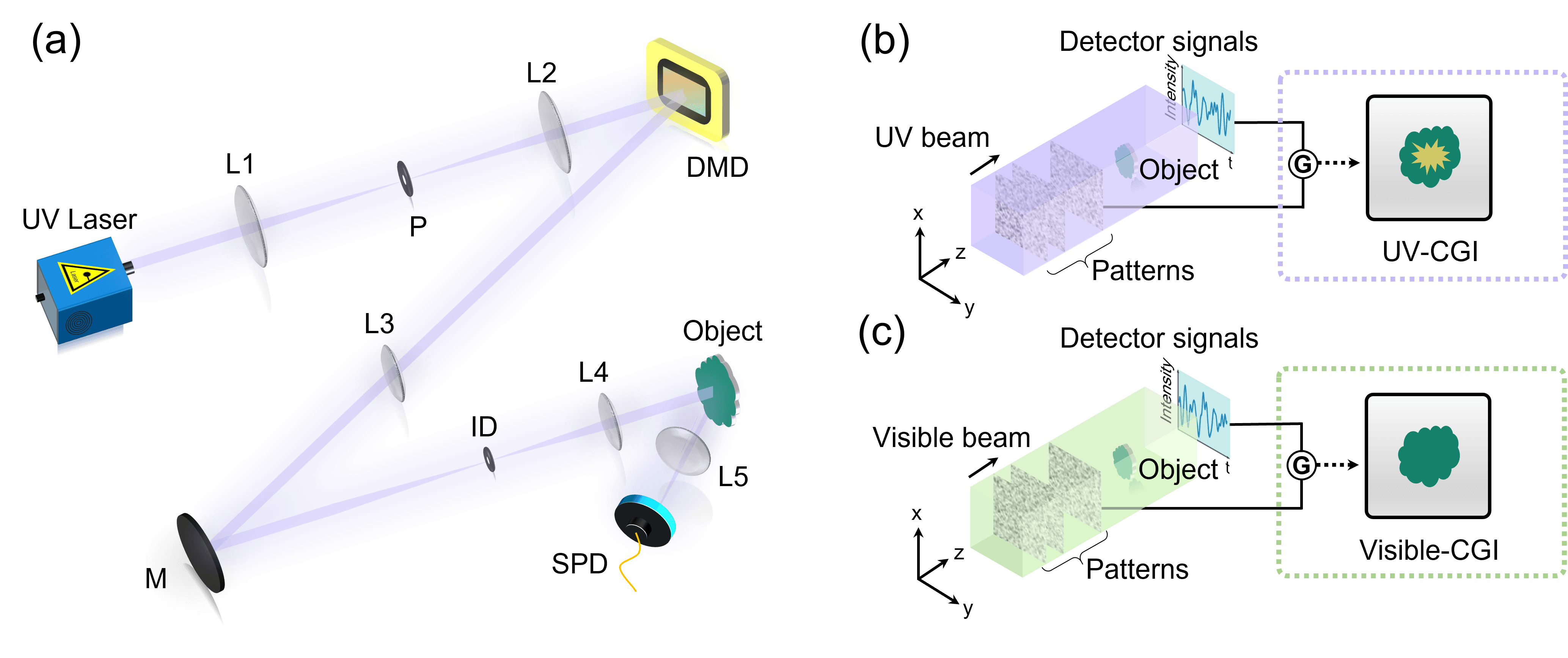}% Here is how to import EPS art
\caption{\label{fig:wide}{(a) Experimental setup. L1-L5: lenses; P: pinhole; DMD: digital micromirror device; M: mirror; ID: iris diaphragm; SPD: single-pixel detector. (b) The UV-sensitive image of the object; the yellow part within the image represents the area that contains UV-sensitive samples. (c) As a comparison, traditional visible CGI cannot reveal such a UV image.}}
\label{fig:1}
\end{figure*}
\section{Experiment and Results}
\subsection{Experimental setup}

The experimental setup of our UV-CGI system is shown in Fig. 1(a). A commercial 325 nm He-Cd laser (IK3301R-G, Kimmon Koha) is used as the light source. The laser beam is first expanded and spatially filtered by a 50 \textmu m pinhole. Then the beam illuminates a high-speed DMD (GmbH V-7001, ViALUX), which generates a series of random modulation patterns at a refresh rate of 10.6 kHz. The patterns have a resolution of $64\times64$ macro pixels ($12\times12$ micromirrors per macro pixel) extending over the central $768\times768$ pixels of the DMD. A $4f$ system consisting of two lenses (L3 and L4) is used to project these patterns onto the object plane. An iris diaphragm (ID) is used to transmit only the zero-order diffraction beam. A collection lens focuses the backscattered light from the sample onto a photodiode SPD (PDAPC2, Thorlabs) which has a spectral response from 320 to 1100 nm. A data acquisition module (USB-6353, National Instruments) synchronizes the DMD and SPD. Samples sensitive to UV light but not visible to the naked eye were chosen as objects. By correlating the illuminating patterns and measured light intensities, the UV-CGI images can be reconstructed by various algorithms, as portrayed in the yellow area of Fig. 1(b), whereas with visible light no image can be seen, as shown in Fig. 1(c). Here, we use the alternating projection algorithm to reconstruct the images \cite{guo2016multilayer}.

\subsection{Demonstration of UV-CGI with UV-sensitive samples}
\begin{figure*}[!t]
\centering
\centering\includegraphics[width=15cm]{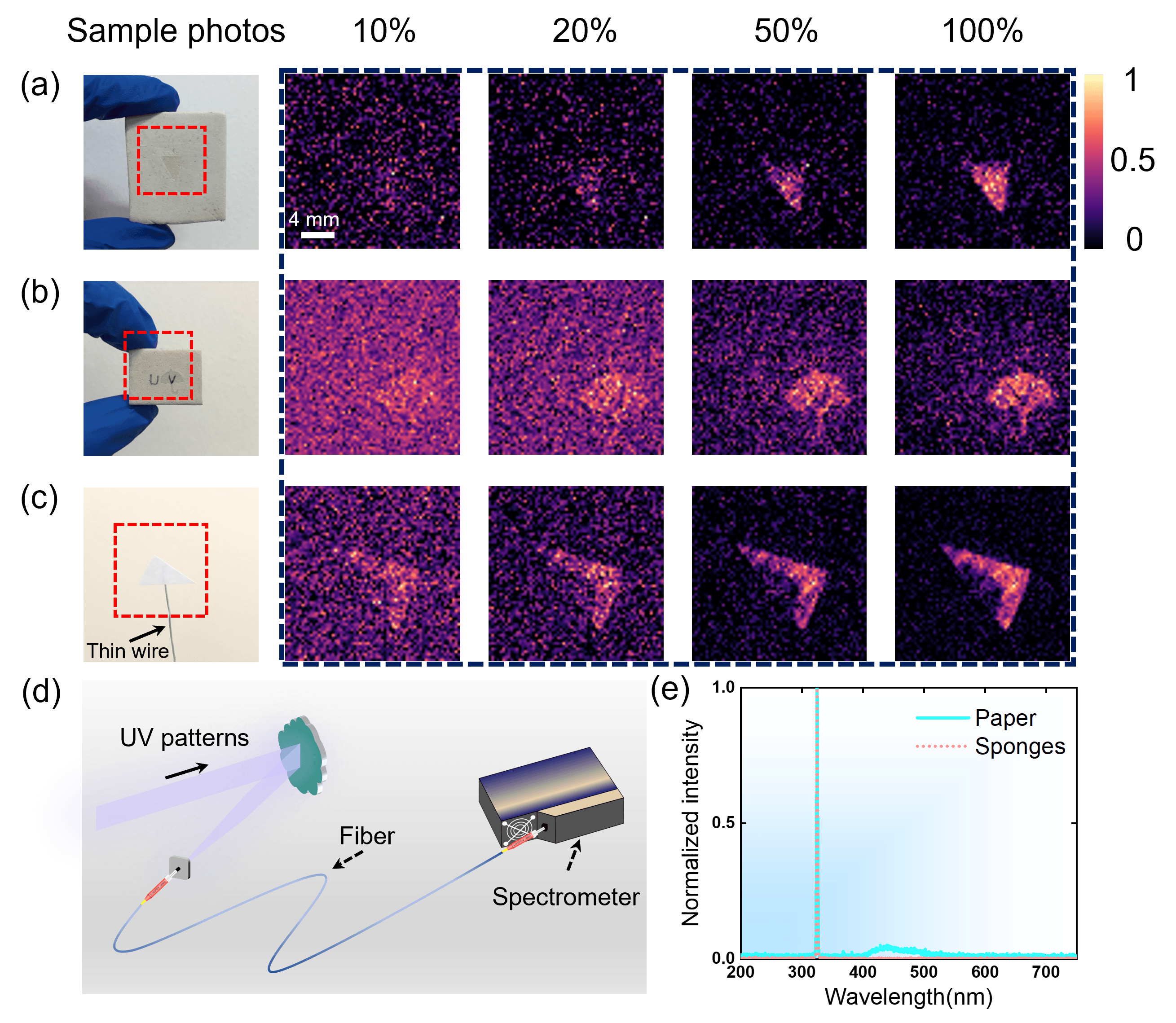}
\caption{\label{fig:wide}{The reconstructed images of three different samples for sampling ratios of 10, 20, 50, and 100$\%$ The left columns show the ground-truth photos of the samples: (a) A sponge coated with sunscreen except in the central triangular area; (b) A sponge coated with sunscreen except in an umbrella-shaped area with the letters “UV” written on it; (c) The bottom part of a triangular piece of paper coated with sunscreen. (d) Schematic diagram of the spectral measurement. (e) The measurement spectra of the paper and the sponges under UV light irradiation.}}
\label{fig:2}
\end{figure*}
We first used chemical sunscreen, which can absorb UV light, as the sample. Three different samples were prepared, as shown in the left columns of Figs. 2(a)-(c). The first was a sponge (38mm$\times$32mm) coated with sunscreen, except for a triangular area in the center. The second one (25mm$\times$18mm) was similar to the first but had an umbrella-shaped uncoated area with a visible "UV" written on it. The area without sunscreen forms an umbrella, as shown in Fig. 2(b). The third sample was a triangular piece of paper supported on a thin wire; a small area near the hypotenuse of the triangle was coated with sunscreen. The reconstructed images ($64\times64$ resolution) of these samples, under different sampling rates, are shown within the black dashed box of Figs. 2(a)-(c). The shapes of the uncoated areas, which cannot be observed clearly by visible photography, emerge with improved clarity as the sampling ratio increases from 10$\%$ to 100$\%$ under UV-CGI. For the first two samples, the triangular and umbrella areas become apparent starting from a sampling ratio of 50$\%$, while the basic shape of the third sample appears at a sampling ratio of 10$\%$. In the second sample the “UV” letters can be seen clearly under visible photography, but the UV-CGI reconstructed image only reveals the shape of the umbrella area, showing the difference between visible and UV photography. For the third sample, the area coated with sunscreen cannot be observed under visible light, as shown in the left column of Fig. 2(c). However, a dark area near hypotenuse of of the triangle is shown clearly in the reconstructed UV images. These results prove that UV-CGI can be used to detect the areas that contain UV-sensitive matter. It should be noted that the reconstructed images of the third sample have better quality than those of the other two samples, especially at a sampling rate of 10$\%$. This is because the third sample has almost no background noise, while the other two suffer from spurious reflection from the area coated with sunscreen. We further measured the spectra of these samples under UV light irradiation, as shown in Fig. 2(d). It can be seen in Fig. 2(e) that all the samples have peaks at 325 nm. However, the sponges barely respond at other wavelengths, showing only a UV spectral line. A small peak is displayed at 400-500 nm for the paper, indicating that the SPD has also collected the visible fluorescence induced by the fluorescent powder in the paper. This is another reason why the reconstructed images of the third sample have better quality than those of the other two samples. 

\begin{figure*}[!t]
\centering
\centering\includegraphics[width=15cm]{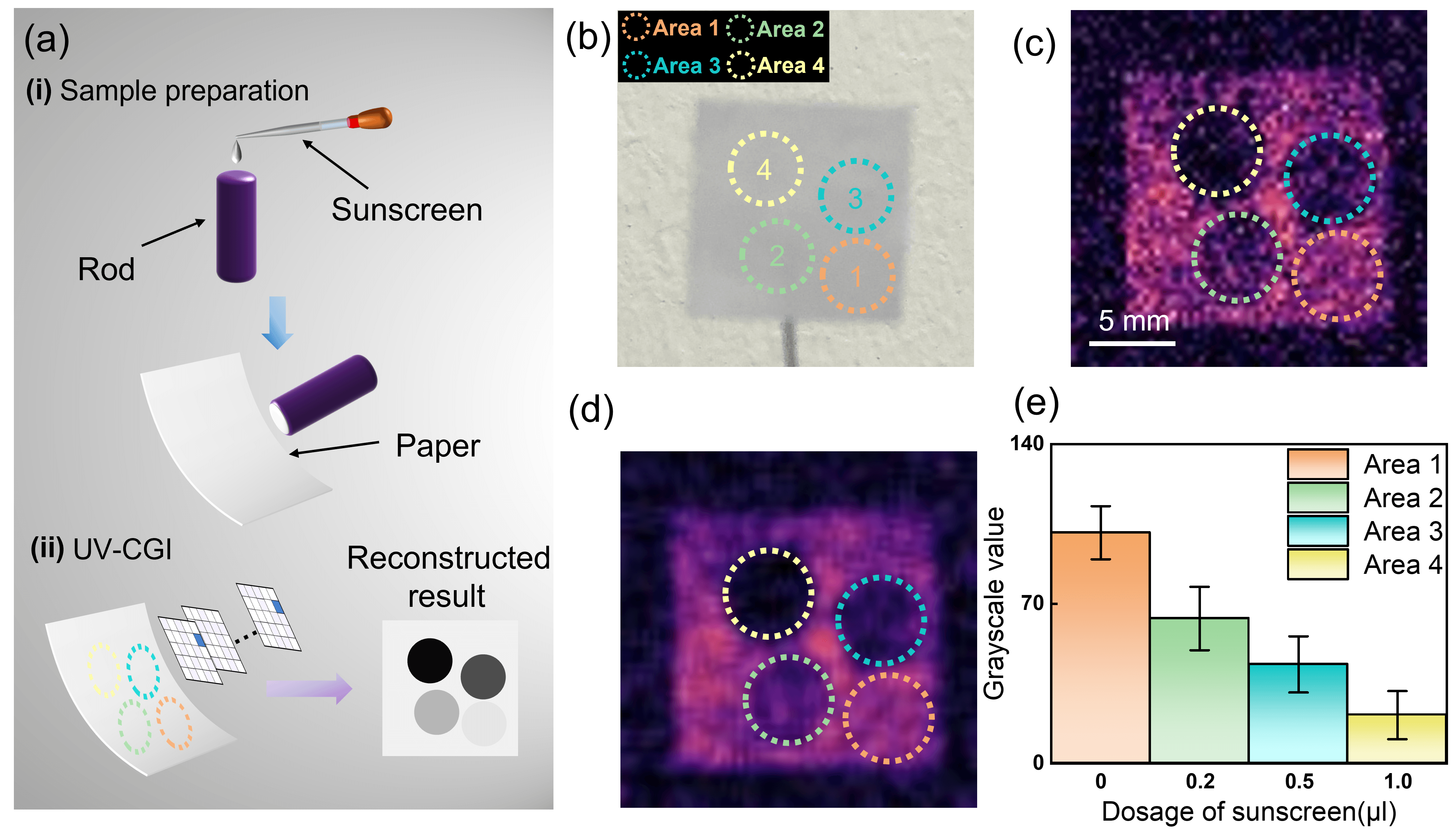}
\caption{\label{fig:wide}{Quantitative detection of sunscreen by UV-CGI. (a) Experimental procedure. (i) Sample preparation.  (ii) UV-CGI exposures. (b) Visible light photos of the sample; the different colors indicate the four areas of the sample coated with different amounts of sunscreen, areas 1-4 containing, respectively, 0, 0.2, 0.5 and 1 \textmu l. (c) Reconstructed image of the sample, in which the four areas are shown with different grayscale values. (d) The smoothed result of (c). (e) Statistical grayscale values of the four areas in (d).}}
\label{fig:3}
\end{figure*}

We further demonstrate that UV-CGI can be used to detect UV-sensitive samples quantitatively. We prepared a sample consisting of four circular areas containing different doses of sunscreen. Figure 3(a) shows a schematic of the sample preparation. We first use a small rod to imprint given doses of sunscreen onto a square piece of paper. A photograph of the sample is shown in Fig. 3(b), in which the differently colored dotted circles indicate the four areas stamped, respectively, with doses of 0, 0.2, 0.5 and 1.0 \textmu l. It is not easy to distinguish the different doses with the naked eye. Figure 3(c) shows the reconstructed UV-CGI images ($64\times64$ resolution) of these four areas from 8192 sampling frames. Different grayscale values are shown each area, corresponding to different densities of the sunscreen. A $3\times3$ kernel is used to smooth the image, as shown in Fig. 3(d). To quantitatively measure the densities of the sunscreen in the four areas, we calculate the standard deviation of the grayscale values, as shown in Fig. 3(e). The statistical results are, respectively, $101.2\pm11.6$, $63.6\pm13.9$, $43.4\pm12.3$, and $21.2\pm10.6$. It can be seen that the higher the sunscreen dosage, the lower the grayscale value. This is because sunscreen absorbs UV light, which decreases the reflected UV light collected by the detector.

\subsection{Dark field UV-CGI to detect phase objects and small defects}
Due to its shorter wavelength, UV reflectance photography has the advantage of detecting fine scratches and defects on metal and semiconductor surfaces, since short wavelength light is more strongly scattered \cite{40ren2022state}. On another note, dark-field imaging or detection is another powerful technique to enhance the contrast of transparent or translucent samples that are not imaged well under bright-field illumination \cite{42gao2021dark}. Instead of collecting the directly transmitted light, dark-field imaging only collects the scattered light from samples, preventing the directly illuminated light from entering the detector. Therefore, it is suitable for revealing sample edges and refractive index gradients, and for the characterization of nanomaterials through measuring the scattering spectra of individual nanoparticles, monitoring dynamic reactions at the single particle level, and so on. 

In this section, we demonstrate a dark-field scheme for UV-CGI to detect the edge of phase objects and small scratches on a CD. The images are reconstructed by measuring the correlation between the intensity values of the scattered field and the DMD patterns. 

\begin{figure*}[!!t]
\centering
\centering\includegraphics[width=15cm]{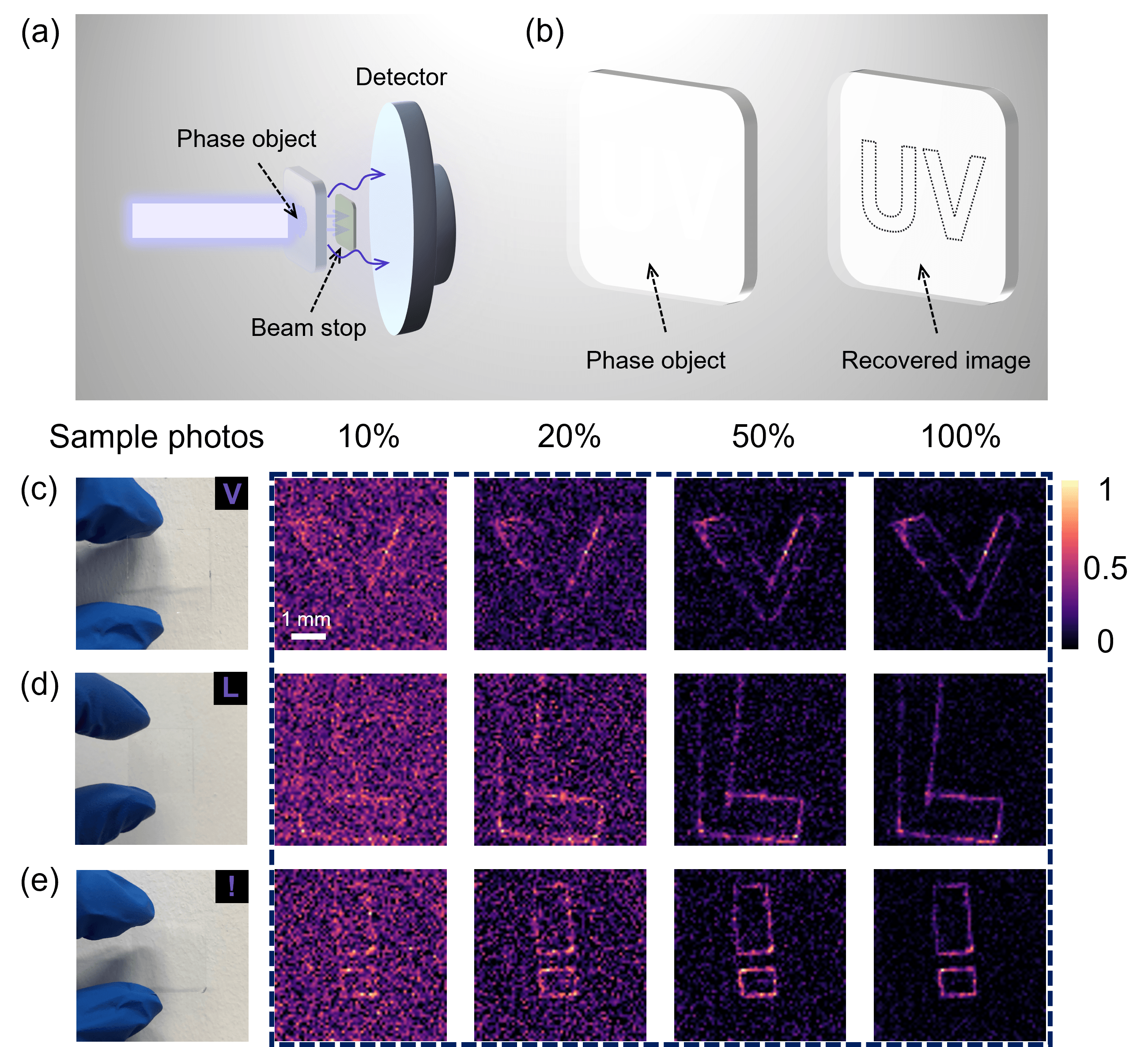}
\caption{\label{fig:wide}{Dark-field UV-CGI for imaging of transmissive pure phase objects. (a) Experimental setup. (b) Schematic of phase object and recovered ghost image. (c)-(e) Reconstructed GI images of three samples for sampling ratios of 10, 20, 50, and 100$\%$. The left columns show the photos of the phase objects, which are the letters: (c) “V”, (d) “L”, and (e) an exclamation mark “!”, respectively.}}
\label{fig:4}
\end{figure*}

\begin{figure*}[!!t]
\centering
\centering\includegraphics[width=15cm]{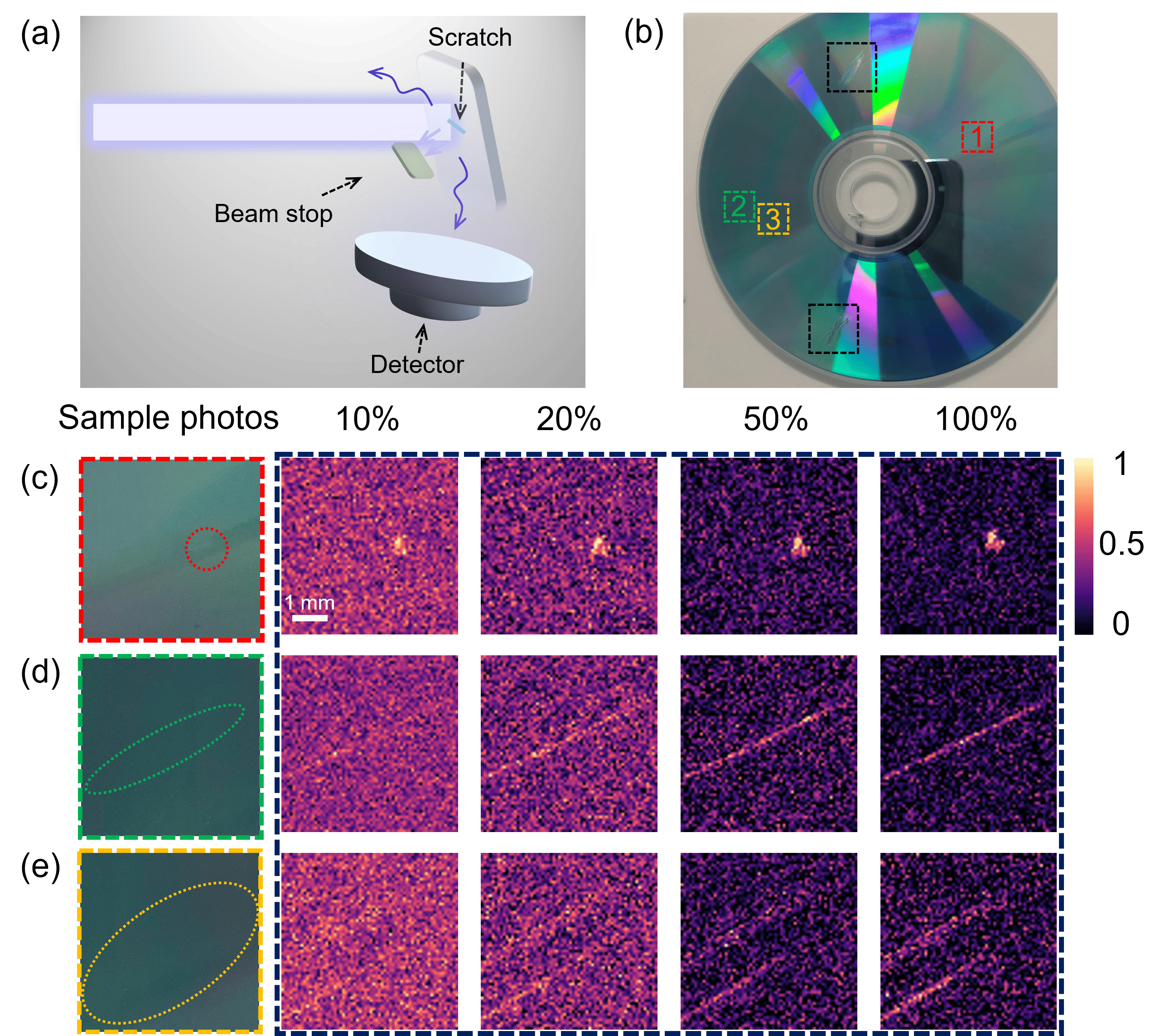}
\caption{\label{fig:wide}{Detection of the damage on a CD by dark-field reflective UV-CGI. (a) Experimental setup. (b) Photograph of the damaged CD. Three slightly damaged areas are shown in the colored boxes, corresponding to: (1) a point-like scratch, (2) a line scratch, and (3) two lines scratches. Another two heavily damaged areas are indicated by the black boxes as a comparison. (c)-(e) Reconstructed images of areas (1)-(3) in (b) for different sampling ratios. The left columns show the magnified photos of the three areas.}}
\label{fig:5}
\end{figure*}

\subsubsection{Dark field UV-CGI with pure phase objects}
Here we show that high-contrast edge images of pure phase objects can be obtained with dark-field UV-CGI. Figure 4(a) shows the experimental setup. The UV beam, after modulation by the DMD, illuminates the object and is then collected by the SPD. A beam stop is placed behind the object to block the direct transmitted light. Figure 4(b) is a schematic diagram of a pure phase object, the edges of which can be revealed after using dark-field UV-CGI. The DMD memory buffer was initially loaded with 4096 patterns with a resolution of $64\times64$ using $192\times192$ DMD pixels. One pixel of the illumination patterns was corresponding to about 82 microns on the object plane. The beam stop is 5.5mm$\times$5.5mm in size, large enough to block all the directly transmitted light. The object was placed just in front about 2 mm from the SPD. Three samples were made from scotch tape and pasted onto a silica substrate. The left column of Figs. 4(c)-(e) shows the photos of the three samples: the letters ‘V’ and ‘L’, and an exclamation mark ‘!’. The sample cannot be seen clearly due to the almost complete transparency of the phase object. The images were reconstructed with sampling ratios of 10, 20, 50, and 100$\%$, as shown in the black dashed box of Fig. 4. The edges of the three samples begin to appear in the image at a sampling ratio of 10$\%$, and become quite clear at a sampling ratio of 50$\%$. The inner part within the edges of the sample is invisible because almost no light is scattered there.

\subsubsection{Detecting damage on a CD}
Figure (5) illustrates the detection of small scratches on a CD by reflective dark-field UV-CGI. Figure 5(a) shows the experimental setup; the directly reflected light from the sample is blocked by the beam block, so the SPD only collects part of the backscattered light from the object. As shown in Fig. 5(b), there are two obvious defects (in the black dashed boxes) and three small scratches on the CD, which are: a point-like scratch inside the red dashed box (1), a scratch inside the green box (2), and two scratches in the orange box (3), respectively. An enlargement of the minor defects is shown in the left column photos of Figs. 5(c)-(e). The DMD patterns are the same as above. The reconstructed images are enclosed in the black dashed box of Figs. 5(c)-(e). It can be seen that the image of the point-like scratch is already visible at the sampling rate of 10$\%$, and is more apparent than the other two line-scratches (Figs. 5(d)-(e)). This is because it is composed of multiple small scratches with more unevenness on the surface so it scatters more light. The reconstructed image in Fig. 5(c) has relatively high contrast at a sampling rate of 50$\%$, almost the same as that at a sampling rate of 100$\%$. However, in Figs. 5(d) and 5(e), the scratches are relatively shallow so scattering is not so strong and the reconstructed images at a sampling rate of 20$\%$ have a highly noisy background. The image quality and contrast improve when the sampling rate increases to 50$\%$ and 100$\%$, especially for the pair of line scratches in Fig. 5(e). 

\section{Conclusion}
In summary, we have experimentally demonstrated UV-CGI using random spatial modulation patterns and a single-pixel detector. Through a quantitative comparison of different densities of UV-sensitive sunscreen samples different grayscale values were obtained for the reconstructed images, representing different sample densities. We have also demonstrated a dark field UV-CGI method to detect phase objects, by collecting only the scattered light from the edges. Due to the short wavelength of UV light, high-contrast images of the fine defects on a CD can be reconstructed, with the capability to resolve $\sim 82$ microns.

Further modifications of UV-CGI can be studied to improve its performance. For example, the system can be optimized with custom designed coded masks or diffusers as the spatial light modulator \cite{17chen2009lensless,22hahamovich2021single,23jiang2021single} to increase imaging speed and reduce system size. Versatile imaging techniques, such as multispectral and time-resolved imaging \cite{44rousset2018time}, can also be introduced for UV-CGI in non-destructive testing, component analysis, or biological fluorescence imaging. Machine-learning algorithms can also help improve imaging speed and the resulting image quality \cite{45he2018ghost}. Our scheme provides a cost-effective UV imaging and detection solution that can open up new opportunities and applications for computational imaging in the UV band.

\bigskip

\begin{acknowledgments}
Funding: National Natural Science Foundation of China (62075004, 11804018, 62275010); China Postdoctoral Science Foundation (2022M720347, 2022TQ0020); Beijing Municipal Natural Science Foundation (4212051,1232027); the Fundamental Research Funds for the Central Universities.
\end{acknowledgments}

\nocite{*}

\bibliography{ref}% Produces the bibliography via BibTeX.

\end{document}